\documentclass[12pt,a4paper,twoside]{article}
\usepackage{epsfig}
\usepackage{baltlat1}
\usepackage{wrapfig}
\begin{document}
\ \
\vspace{0.5mm}

\setcounter{page}{1}
\vspace{8mm}

\titlehead{Baltic Astronomy, vol.12, XXX--XXX, 2003.}

\titleb{GIANT RADIO SOURCES IN VIEW OF THE DYNA-\\MICAL EVOLUTION
OF FRII-TYPE POPULATION}

\begin{authorl}
\authorb{J.~Machalski}{1} 
\authorb{K.~Chy\.{z}y}{1} and
\authorb{M.~Jamrozy}{2}

\end{authorl}

\begin{addressl}
\addressb{1}{Astronomical Observatory, Jagellonian University,
ul. Orla 171,\\ 30-244 Cracow, Poland}

\addressb{2}{Radio Astronomical Institute, Bonn University, 
Auf dem H\"ugel 71,\\ 53-121 Bonn, Germany }
\end{addressl}

\submitb{Received: xx October 2003}

\begin{abstract}
The time evolution of {\sl giant} ({\footnotesize $ D>1$}\, Mpc) 
lobe-dominated galaxies is analysed on the basis of dynamical 
evolution of the entire FRII-type population.
\end{abstract}

\resthead{Dynamical Evolution of Giant Radio Sources}
{J.~Machalski, K.~Chy\.{z}y and M.~Jamrozy}



\sectionb{1}{INTRODUCTION}
One of the general questions concerning the largest radio sources is: do 
they reach their extremal {\sl giant} sizes due to (i) exceptional 
physical conditions in the intergalactic medium, (ii) extraordinary 
intrinsic properties of the AGN, or simply (iii) because they are 
extremely old? To answer this question, a number of attempts were 
undertaken to recognize properties other than size which differentiate 
{\sl giants} from normal-size sources. As a result of those attempts, 
a recent notion is that the largest sizes result from a combination
of the above properties.

In this contribution we analyse whether properties of {\sl giant} radio
galaxies observed in a selected representative sample can be explained 
by a model of the dynamical evolution of classical double radio 
sources in cosmic time, and what factor (if there is a one) is
primarily responsible for the {\sl giant} size. Two recent analytical 
models, published by Kaiser et al. (1997) [hereafter KDA] and Blundell 
et al. (1999), are very convenient for this purpose. These two models 
differ in predictions of the time evolution of the source luminosity 
(there is no space in this contribution to go into details). In summary, 
basic physical parameters, i.e. the jet power 
{\footnotesize$Q_{jet}$}, the central density of the galaxy nucleus 
{\footnotesize$\rho_{0}$}, the energy density and pressure in the 
lobes or cocoon ({\footnotesize$u_{c}$} and {\footnotesize$p_{c}$}), 
and the total energy of the source {\footnotesize$E_{tot}$} are 
derived from the model for each member of the sample to fit its age, 
redshift, radio luminosity, projected size, and axial ratio. Next, 
these parameters are compared with (1) the relevant parameters derived 
for normal-size sources in a comparison sample, and (2) the parameters
determined from observational data, i.e. the age, equipartition energy 
density {\footnotesize$u_{eq}$}, equipartition energy 
{\footnotesize$U_{eq}$}, etc., calculated under `minimum energy' 
conditions.

A description of the observational data used, application of the 
analytical model, fitting procedure, etc. (cf. Machalski et al. 2003), 
are beyond the scope of this contribution.

\sectionb{2}{RESULTS OF THE MODELLING}

\noindent
{\it \underline{Jet power {\footnotesize$Q_{jet}$} and core density 
{\footnotesize$\rho_{0}$}}}

\begin{wrapfigure}{i}[0pt]{58mm}
\centerline{\psfig{figure=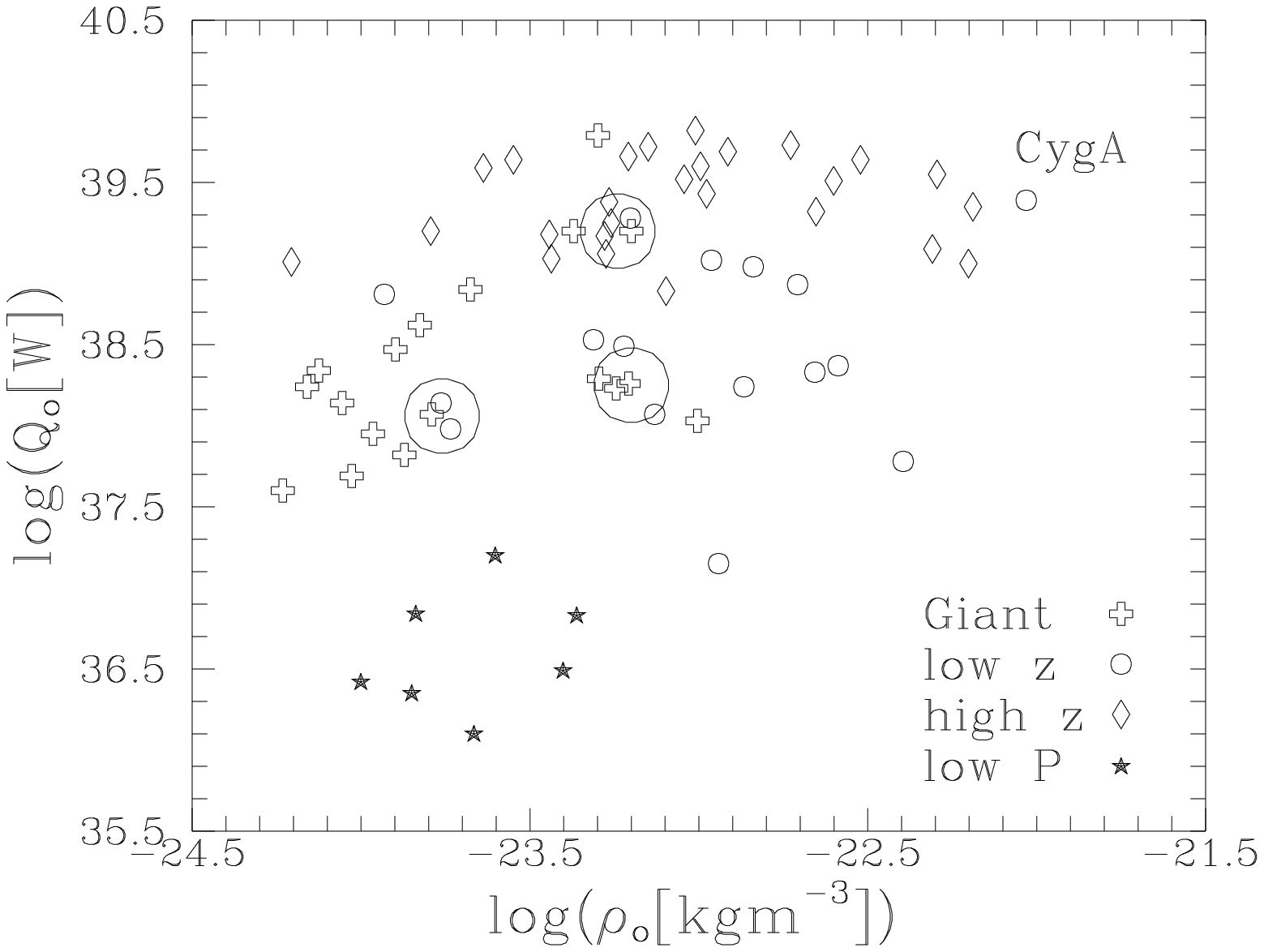,width=65truemm,angle=0,clip=}}
\captionb{1}{{\footnotesize Plot of the jet power 
{\footnotesize$Q_{jet}$} against the core {\footnotesize$\rho_{0}$}.
The `clans' of sources with similar {\footnotesize$Q_{jet}$} 
and {\footnotesize$\rho_{0}$} are marked by the large circles}}
\end{wrapfigure}

A distribution of the above parameters on the
{\footnotesize $\log(Q_{jet}$})--{\footnotesize$\log(\rho_{0}$}) 
plane is shown in Fig.~1. As both parameters should be independent, 
we test whether the observed distribution is or is not biased by 
possible selection effects. The data obtained imply that 
{\footnotesize$Q_{jet}$} correlates, in order of the  significance 
level, with the luminosity {\footnotesize$P_{1.4}$}, redshift 
{\footnotesize$z$} and age {\footnotesize$t$}. Calculating the Pearson
partial correlation coefficients, we found no significant correlation 
between {\footnotesize$Q_{jet}$} and {\footnotesize$\rho_{0}$} when 
{\footnotesize$z$} (or {\footnotesize$P_{1.4}$}) and 
{\footnotesize$t$} are kept constant.

In view of the above, one can see from Fig.~1 that (i) among the 
sources with a comparable {\footnotesize$Q_{jet}$}, {\sl giant} 
sources have an average {\footnotesize$\rho_{0}$} smaller than its 
corresponding value in normal-size sources, (ii) {\sl giants} have at 
least ten times more powerful jets than much smaller low-luminosity 
sources of a comparable {\footnotesize$\rho_{0}$}, and (iii) for a 
number of sources in the sample, the derived values of their 
fundamental parameters {\footnotesize$Q_{jet}$} and 
{\footnotesize$\rho_{0}$} are very close, while their ages and axial 
ratios are evidently different. Thus in view of the model assumptions,
they may be considered as `the same' source observed at different 
epochs of its lifetime. Such bunches of three to five sources 
(hereafter called `clans') are indicated in Fig.~1 with the large 
circles.

\noindent
{\it \underline{Relation between {\footnotesize$D$}, 
{\footnotesize$Q_{jet}$}, {\footnotesize$\rho_{0}$}, 
{\footnotesize$t$}, and {\footnotesize$(1+ z)$}}}

In view of the dynamical model applied and as a result of the Pearson 
partial-correlation coefficients calculated between those parameters, 
we find that the linear size of a source strongly depends on both
its age and the jet power, while the correlation with age is the 
strongest. However, the size also anti-correlates with central 
density of the core. That anticorrelation seems to be weaker than the 
correlations with {\footnotesize$Q_{jet}$} and {\footnotesize$t$}
and become well pronounced only when all three remaining parameters 
({\footnotesize$Q_{jet}$}, {\footnotesize$t$} and {\footnotesize$z$}) 
are kept constant.

\sectionb{3}{EVOLUTIONARY TRACKS OF SOURCES}

In the papers of KDA and Blundell et al. the tracks of radio 
luminosity {\footnotesize$P_{\nu}$}
versus linear size {\footnotesize$D$} were derived for imaginary 
sources with assumed values of {\footnotesize$Q_{jet}$}, 
{\footnotesize$\rho_{0}$}, {\footnotesize$z$} and other free 
parameters of the model. In our approach we are able to calculate 
such evolutionary tracks for \underline{actual} sources. In 
Sect.\,2, the `clans' of sources have been pointed out. Three of six
clans are marked in Fig.~1. Since the dynamical model assumes constant
{\footnotesize$Q_{jet}$} during a source lifetime, and a nucleus 
density {\footnotesize$\rho_{0}$} is {\sl a priori} constant, members 
of such a clan can be considered as `the same' source observed at a 
number of different epochs throughout its life. The observed parameters
of these members can verify predictions of the model. However, fits of 
the {\footnotesize$P$}--{\footnotesize$D$} tracks predicted with the 
original KDA model to the observed parameters of sources have appeared 
unsatisfactory. Much better fits of the modelled tracks to 
observational data of the `clan' members are found \underline{with the 
cocoon's axial ratio ($AR$) evolving in time}. This, in turn, implies a time 
evolution of a ratio of the jet head pressure to the cocoon pressure, 
{\cal P}{\footnotesize$_{hc}$}.

\noindent
{\it \underline{The tracks {\footnotesize$\log
P_{1.4}$}--{\footnotesize$\log D$}
and {\footnotesize$\log u_{eq}$}--{\footnotesize$\log E_{tot}$}}}

\begin{figure}
\centerline{\psfig{figure=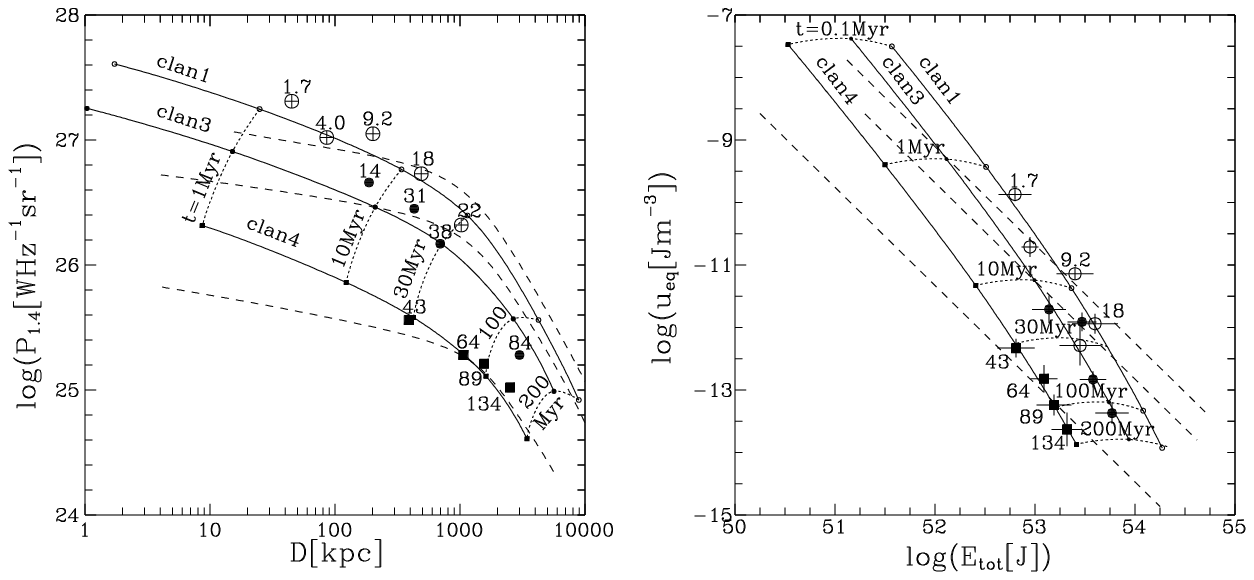,width=130truemm,angle=0,clip=}}
\captionb{2(a)}{{\footnotesize Evolutionary {\footnotesize$P-D$} 
tracks and {\bf (b)} {\footnotesize$u_{eq}-E_{tot}$} tracks fitted for 
three clans of sources with evolving axial ratio {\footnotesize$AR$} 
(solid curves). The markers of the same predicted age on each curve 
are connected with dotted lines. The members of each clan are marked 
with different symbols. Their actual age is indicated by a number 
behind the symbol. The dashed curves indicate relevant tracks but 
calculated with a constant {\footnotesize$AR$}, as in original KDA 
model}}
\end{figure}

The evolutionary {\footnotesize$P-D$} tracks for the three clans are 
shown in Fig.~2a with solid curves. The markers of the same time are 
put on these tracks. The dashed curves show the relevant tracks 
calculated from the original KDA model, i.e. with a constant 
{\footnotesize$AR$} taken as the mean of axial ratios in a given clan. 
It is clearly seen that the evolving {\footnotesize$AR$} much better 
fits the observed changes of {\footnotesize$P$} on {\footnotesize$D$}.

The model also allows a prediction of the source's evolution on the 
energy density--total energy plane which are shown in Fig.~2b. It is 
worth emphasizing that the predicted evolutionary 
{\footnotesize$u_{c}$}--{\footnotesize$E_{tot}$} tracks are 
\underline{steeper} and \underline{curved} in respect to those 
expected from the original KDA model. Moreover, the steepening 
increases throughout the source lifetime. This is caused by the 
non-constant adiabatic losses and inflation of the cocoon in time, as 
well as a faster decrease of the cocoon pressure in very large sources.
 Quantitatively this process is evaluated by a substitution of the 
evolving (increasing) value of the pressure ratio 
{\cal P}{\footnotesize$_{hc}(t)$} into equation describing 
{\footnotesize$E_{tot}$} (cf. Machalski et al. 2003).

ACKNOWLEDGMENTS.\ MJ acknowledges the financial support from EAS.
\goodbreak

References\\
Blundell~K.~M., Rawlings~S., Willott~C.~J. 1999, AJ, 117, 766\\
Kaiser~C.~R., Dennett-Thorpe~J., Alexander~P. 1997, MNRAS, 292, 723 (KDA)\\
Machalski~J., Chy\.{z}y~K.~T., Jamrozy~M. 2003 (submitted for MNRAS, astro-ph/0210546)

\end{document}